# PHASE STATISTICS AND THE HAMILTONIAN

*I.R.Peterson*

Centre for Molecular and Biomolecular Electronics
Coventry University SE
Priory Street
Coventry CV1 5FB ENGLAND

*Abstract:*

Modern statistical thermodynamics retains the concepts employed by Landau of the order parameter and a functional depending on it, now called the Hamiltonian. The present paper investigates the limits of validity for the use of the functional to describe the statistical correlations of a thermodynamic phase, particularly in connection with the experimentally accessible scattering of X-rays, electrons and neutrons. Guggenheim's definition for the functional is applied to a generalized system and the associated paradoxes are analyzed. In agreement with Landau's original hypothesis, it is demonstrated that the minimum is equal to the thermodynamic free energy, requiring no fluctuation correction term. Although the fluctuation amplitude becomes large in the vicinity of a second-order phase transition in low dimensionalities, it does not diverge and the equilibrium order parameter remains well defined.

Email: I.Peterson@Coventry.ac.uk

# 1. INTRODUCTION

Partially ordered phases of organic materials are of considerable importance in biology and technology. They have also been interesting test cases for theories of statistical mechanics. One of the triumphs of renormalization group (RG) techniques was the theoretical prediction of the hexatic phase by Nelson and Halperin [1,2]. Hexatics have since been observed in a wide range of lamellar organic systems [3,4].

The present investigation has been inspired by the possibility of deducing in a rigorous manner the molecular organization of a partially ordered lamellar phase from its experimentally accessible scattering profiles [5]. In particular, in spite of their lack of long-range translational order, hexatic phases give rise to diffraction patterns showing discrete reflections resembling those of a crystal [3,4].

There have been a number of analyses of the diffraction profile of a hexatic phase in the literature [6,7] but up until recently all were phenomenological, some relying on intuitive auxiliary hypotheses, and they provided no access to molecular variables. The situation has now been improved for hexatic smectic B states, whose diffraction profile can now be expressed in terms of microscopically meaningful parameters [8]. The crux of the analysis for these states was the ability to treat a particular discrete variable — the Burgers vector at each lattice site — as if it had a continuous statistical distribution. The assumption of continuity allows sums in the expression for the Patterson, or density-density correlation function, to be converted to integrals so that they can be evaluated.

Many statistical thermodynamic theories also treat the variables of a thermodynamic system as if they were continuous. The continuous variables are known as order parameters. Such theories include Landau's theory of phase transitions [9,10,11] through Kadanoff's scaling laws [12] to Wilson's renormalization group theory [13,14]. All rely on the concept of a thermodynamic quantity depending on the order parameter field, which is known either as the free-energy functional $\Phi$ or as the Hamiltonian $H$. This functional is considered to define the following probability distribution function for the continuous variables:

$$f(\bar{s}) = \mathbf{k}\exp[-H(\bar{s})/kT].  \quad \ldots\ldots\ldots\ldots\ldots\ldots\ldots\ldots\ldots\text{(Eqn 1)}$$

An explicit expression of this sort is exactly what is required to allow evaluation of the scattering profile.

On a number of well-documented occasions, behavior intuited for the free energy functional has proved to be incorrect. For example, shortly after Landau theory was presented, its validity was cast into doubt by Onsager's analysis of the two-dimensional Ising model [15], which involved the exact evaluation of the statistical mechanics of the system. Onsager's computed variation of the specific heat for this particular model was found to be incompatible with Landau's assumption of completely analytic behavior of the free energy functional. Unfortunately Onsager's exact solution has only been extended to a limited number of other systems [16,17] and is not a general approach to understanding thermodynamic systems. However further work by Kadanoff and Wilson refined the idea of fluctuations [18,19]. It eventually led to renormalization group methods, which retain the concepts of the order parameter and an energy functional.

These concepts are not free of difficulties. Levanyuk [20] and Ginzburg [21] analyzed the effect of order-parameters fluctuations by assuming, in analogy with other formalisms, that the integral of this density was not the thermodynamically valid value for the free energy, but had to be corrected by a term expressing the effect of fluctuations. They were able analytically to evaluate their integral expression for the fluctuation correction. In the vicinity of a second-order phase transition of a two-dimensional system, the revised theory agrees with the Onsager analysis in predicting a logarithmic discontinuity in the specific heat. According to their analysis, the non-analytic behavior at these phase transitions results from long-wavelength fluctuations. By implication, the Hamiltonian obtained by subtracting the fluctuation correction term from the true free energy gives a short-wavelength "core" energy density, which is an analytic function of the order parameter and therefore has most of the properties of the free-energy functional in the original Landau theory. Unfortunately this theory trades one problem for another, because the amplitude of the fluctuations of each order parameter in a system of low dimensionality diverges in the vicinity of a second-order phase transition.

In spite of these difficulties, Landau theories are often found to provide a good description of experimental phase transitions [22]. The discrepancies are expected to be most severe for low-dimensional systems, yet the Landau theory for the two-dimensional lamellar phases of rod-like (i.e. one-dimensional) molecules [23,24] is in excellent agreement with results from Brewster-angle and fluorescence microscopy complemented by grazing-incidence diffraction. It is tempting to use the free-energy functional proposed in these papers to describe the molecular statistics in these systems, but for this to be convincing, this procedure must first be demonstrated to be valid.

Guggenheim is one of the few authors to have defined a suitable functional [25]. In renormalization group analyses, the definition of the Hamiltonian is usually glossed over, and the analysis starts with a functional in which the order parameters are already averaged over many molecular spacings. The present paper fills a gap in the literature by refining and analyzing his formalism to check the associated paradoxes. The divergence of the fluctuation amplitude is equivalent to the assumption that the probability density is given by Eqn 1. This statistical distribution, with $\Phi$ equal to the Guggenheim free energy functional, correctly gives the moments of the order parameter to second order, but no higher. The true fluctuation amplitude is seen to remain finite even at a second-order phase transition. The fluctuation contribution to the free energy is investigated. The Guggenheim definition is shown to include the effects of fluctuations, in the sense that its minimum is equal to the system free energy, requiring no correction term.

## 2. DEFINITION

It is well known that the Helmholtz free energy $F$ of classical thermodynamics in is defined in terms of the Gibbs [26] partition function $Z$ in a quantum context by:

$$\exp[-F/kT] = Z$$
$$\equiv \sum_r \exp[-E_r/kT] \qquad \ldots\ldots\ldots \text{(Eqn 2)}$$

where the sum is taken over all quantum states $r$ of the system. All the other classical thermodynamic functions of internal energy, entropy, enthalpy, Gibbs free energy, etc., can be expressed in terms of $F$ and its derivatives, so that a knowledge of $Z$ completely determines the classical thermodynamics of the system.

Eqn 2 gives the value of the free energy applicable for the state of thermodynamic equilibrium. In contrast, the functional defined by Guggenheim is not limited to equilibrium, but depends on an order parameter. It is defined by an equation very similar to Eqn 2, but with Z replaced by an incomplete partition function $Z(\boldsymbol{x})$ in which the sum is restricted to states in which selected measurable parameters are equal to given values $\boldsymbol{x}$.

Statistical thermodynamic parameters can only be evaluated directly for finite systems; for an infinite system the free energy per unit volume or mass must be derived as the limit of the series of values calculated for finite systems. Finite systems have only a countably infinite number of quantum states, so that most values of $\boldsymbol{x}$ do not correspond to any states. To achieve a continuous function, this definition must be further refined in terms of the Gibbsian ensemble, which is a set of $N$ replicas of the given finite system, and allowing $N$ to increase without limit [27]. Even then, when $N$ is finite, the states of the ensemble are still countable, so that it is necessary to include in the sum all states in some finite neighborhood of the given values $\boldsymbol{x}$. As $N$ tends to infinity, the neighborhood is then allowed to shrink to zero according to a power law following the standard procedure[28].

Explicitly in the present notation, the refinement of the definition of Guggenheim involves an ensemble of $N$ non-interacting replicas of the system. The number $R$ of states of each replica is constrained to be finite, and the $r$-th state is labeled by an element $s_r$ of a given, but arbitrary, vector space. $E_r$ is the internal energy of the $r$-th state, and the probability that the replica is in the $r$-th state is given by $k.\exp(-E_r/kT)$, where the prefactor $k$ is chosen to give unity sum. A state $\mathbf{r}$ of the ensemble involves the first replica being in state $r = r_1$, the second in state $r_2$, etc.; its energy $E_{\mathbf{r}}$ is simply the sum of the individual energies; and its average order parameter $\langle s \rangle_{\mathbf{r}}$ is the mean of the individual structural variables. The free energy functional for the complex system involves the incomplete partition function over states in the neighborhood of a given value $\bar{s}$:

$$\frac{\Phi(\bar{s})}{kT} = -\lim_{N\to\infty}\left\{N^{-1}.\ln\left(\sum_{|\langle s\rangle_{\mathbf{r}} - \bar{s}| < N^{-d}} \exp\left[-\frac{E_{\mathbf{r}}}{kT}\right]\right)\right\}$$

.................... (Eqn 3)

The limit is independent [28] of the neighborhood exponent $d$, most conveniently chosen in the range $0 < d < 0.5$. The mathematical refinement of Eqn 3 clearly retains the spirit of the Guggenheim definition [25], as well as the notion that the order parameter is the average of a physical system parameter (in the present case spatial averaging is ruled out).

In Section 3 an explicit expression will be derived for the free-energy functional defined in Eqn 3, and its absolute minimum value deduced. For the probability analysis of Section 4 it is necessary to restrict $E_r$ to a bilinear form in $s_r$ with matrix $V$

$$E_r = s_r^T V s_r \qquad \text{.................... (Eqn 4)}$$

This formalism covers many interesting systems, including the Ising magnetic transition, its generalization by Potts, lattice gases, and the Kosterlitz-Thouless-Nelson-Halperin-Young defective crystal.

## 3. EXPLICIT DERIVATION AND MINIMIZATION

Consider the average $\langle s \rangle$ of the vectors for the Gibbs ensemble of $N$ replicas of the system. Suppose a particular value $\bar{s}$ of the mean is achieved with fractions $p_r$ of each of the $R$ possible vectors, respectively:

$$\bar{s} = \sum_{r=1}^{R} p_r s_r \quad \text{.........................(Eqn 5)}$$

The sum of all fractions must equal unity:

$$\sum_{r=1}^{R} p_r = 1 \quad \text{.........................(Eqn 6)}$$

The logarithm of the number of combinations where the fractions differ by less than $N^{-d}$ from $p_r$ is given by:

$$\ln\binom{N}{Np_1 \; \text{-} \; Np_r \; \text{-} \; Np_R} = -N\sum_r p_r \ln p_r - O(\ln N)$$

$$\text{.........................(Eqn 7)}$$

Taking into account the energy factor $\exp(-E_r/kT)$, the negative logarithm of the corresponding contribution to the partition function is given to sufficient accuracy by:

$$\sum_r \{p_r \ln p_r + p_r E_r/kT\} \qquad \ldots\ldots\ldots\ldots\ldots\ldots\text{(Eqn 8)}$$

Now the function $p_r \ln p_r$ has everywhere a positive second derivative within its physically meaningful range (0,1). Hence over the (R-2)-dimensional subspace defined by Eqns 6 and 7, the function of Eqn 8 has a unique minimum, with a width in all directions of the order of $N^{-1/2}$. In the limit of infinite ensemble size, the absolute minimum of Eqn 8 within this subspace gives the Guggenheim free energy functional.

Using the method of Lagrange multipliers, the values of the fractions $p_r$ at this minimum are determined to be:

$$p_r = \frac{\exp(\mathbf{h}\cdot s_r - E_r/kT)}{\sum_r \exp(\mathbf{h}\cdot s_r - E_r/kT)} \qquad \ldots\ldots\ldots\ldots\ldots\ldots\text{(Eqn 9)}$$

The vector $\mathbf{h}$ is the undetermined multiplier. The denominator is the moment generating function of $s$ considered as a random variable. It will be used to define an auxiliary quantity $\Gamma$ with the dimensions of energy:

$$\exp[-\Gamma(\mathbf{h})/kT] \equiv \sum_r \exp[\mathbf{h}\cdot s_r - E_r/kT]$$

$$\qquad\qquad\qquad\qquad\qquad \ldots\ldots\ldots\ldots\ldots\ldots\text{(Eqn 10)}$$

The value of $\mathbf{h}$ must be determined by simultaneous solution with Eqn 10, i.e.:

$$\bar{s} = \frac{\sum_r s_r \exp(\mathbf{h} \cdot s_r - E_r/kT)}{\sum_r \exp(\mathbf{h} \cdot s_r - E_r/kT)}$$

$$= -\nabla \Gamma(\mathbf{h})/kT$$

................................. (Eqn 11)

It can be shown [29] that this relationship determines $\mathbf{h}$ uniquely once $\bar{s}$ is given. For a given value of $\bar{s}$, Eqn 9 then gives the weights $p_r$ which can be substituted into Eqn 8. The normalized free energy simplifies to:

$$\Phi(\bar{s}) = \Gamma(\mathbf{h}) + kT\, \bar{s} \cdot \mathbf{h} \qquad \ldots\ldots\ldots\ldots \text{(Eqn 12)}$$

In the light of Eqn 11, Eqn 12 is seen to be a Legendre transform of the function $\Gamma$. Hence the unique inverse of Eqn 12 can be expressed explicitly in terms of $\Phi$ by:

$$\mathbf{h} = \nabla \Phi(\bar{s})/kT \qquad \ldots\ldots\ldots\ldots\ldots\ldots \text{(Eqn 13)}$$

Eqn 13 shows immediately that the minimum value of $\Phi$ always occurs for zero $\mathbf{h}$. Making this substitution in Eqns 12 and 10, the resultant minimum value of the Guggenheim free energy functional $\Phi$ is seen to be identical to the free energy $F$ defined in Eqn 2, as stated by Landau [9].

It can be seen that the initial step in computing the free energy functional is to determine the auxiliary function $\Gamma(\mathbf{h})$ as defined in Eqn 10. This has exactly the same form as the free energy defined in Eqn 2, but greater complexity due to the presence of the fictitious field $h$. Just as $F$ is not necessarily an analytic function of its arguments, then neither need $\Gamma$ be. The calculation may not be straightforward and it may only be possible to obtain an approximation using, for example, renormalization group methods. However the resulting functional provides information about microscopic phase statistics via Eqns 1 and 12. The next Section will investigate the validity of Eqn 1 in the light of the known paradoxes.

## 4. PROBABILITY DISTRIBUTION FUNCTION

Eqn 1 is commonly employed to describe the probability distribution of fluctuations, e.g. Guggenheim [25] and Ma [13]. In this Section, its validity will be investigated. Since the present system is discrete, whereas the probability distribution of Eqn 1 is inherently continuous, it is necessary to assume that the number of states R of the system is large, with a density of states $n(s)$ in the vicinity of any given value of the state vector. Note that the argument here does not involve averaging over the ensemble, so that the symbol $s$ is used rather than $\bar{s}$.

Within the restrictions of the present system, the exact probability distribution function is given by:

$$f(s) = \mathbf{k} n(s) . \exp\left(-\frac{s^T V s}{kT}\right) \quad \text{(Eqn 14)}$$

Eqns 10 and 12 are readily applied to derive power series expressions for the functions $\Gamma$ and $\Phi$ correct to second order in their respective arguments:

$$\Gamma(\mathbf{h}) = \langle s \rangle \mathbf{h} + \tfrac{1}{2} C \mathbf{h}^2$$

$$\Phi(s) = \tfrac{1}{2}(s - \langle s \rangle)^T C^{-1}(s - \langle s \rangle)$$

(Eqn 15)

where the mean $\langle s \rangle$ and the correlation tensor $C$ are defined by:

$$\langle s \rangle = \frac{\sum_r s_r \exp(-E_r/kT)}{\sum_r \exp(-E_r/kT)}$$

$$C = \frac{\sum_r s_r^2 \exp(-E_r/kT)}{\sum_r \exp(-E_r/kT)} - \langle s \rangle^2$$

................ (Eqn 16)

Clearly, this degree of approximation always leads to a multivariate Gaussian distribution. At the same degree of approximation, the density of states may also be taken to be a Gaussian:

$$n(s) = \exp(L^T s - s^T M s) \quad \ldots\ldots\ldots\ldots\ldots \text{(Eqn 17)}$$

from whence the mean and correlation tensors are evaluated to be:

$$C = \left(M + \frac{V}{kT}\right)^{-1}$$

$$\langle s \rangle = \tfrac{1}{2} C L$$

......................... (Eqn 18)

Hence Eqn 1 is correct to this approximation, which is appropriate when experimental measurements involve linear combinations of very many individual order parameters. Expanding the free energy functional to higher order terms involves a knowledge of the higher-order moments $\langle s^3 \rangle$, etc. In this case the Central Limit Theorem ensures that none of these can be observed.

The accuracy of the fluctuation statistics to higher moments is most readily checked by the following argument in which, for simplicity, $kT$ is taken to be unity. Suppose that the true log moment generating function is given, correct to fourth order, by:

$$\Gamma(\mathbf{h}) = -\tfrac{1}{2} A \mathbf{h}^2 + \tfrac{1}{4} B \mathbf{h}^4 \quad \ldots\ldots\ldots\ldots\ldots \text{(Eqn 19)}$$

The tensors $A$ and $B$ can be taken without loss of generality to be symmetrical, so that this compact notation is unambiguous. Then the free energy functional is given, also correct to fourth order, by:

$$\Phi(\bar{s}) = \tfrac{1}{2}A^{-1}\bar{s}^2 + \tfrac{1}{4}BA^{-4}\bar{s}^4 \qquad \ldots\ldots\ldots\ldots\ldots (\text{Eqn 20})$$

Consider the random variable whose probability distribution function is $k.\exp[-\Phi(s)]$. Its log moment generating function is readily evaluated to be, correct to fourth order in $\mathbf{h}$ and to first order in $B$:

$$\ln\langle e^{\mathbf{h}\cdot s}\rangle = -\tfrac{1}{2}(A + 6BA^{-1})\mathbf{h}^2 + \tfrac{1}{4}B\mathbf{h}^4. \qquad \ldots\ldots\ldots\ldots\ldots (\text{Eqn 21})$$

Hence, whilst Eqn 1 is valid at a useful level of approximation, its applicability extends no further.

The validity of Eqn 1 can also be checked by examining specific cases. Perhaps the simplest is that of an isolated spin that can take on two orientations, both with the same internal energy. Figure 1 shows three related functions: the true probability distribution function, the equivalent Gaussian, and Eqn 1. In this very simple case, there is no doubt that all three are distinct functions.

The paradox involving the divergence of the order parameter fluctuations in low dimensionalities at a second-order transition arises when the smallest eigenvalue of the internal energy matrix $V$ of Eqn 4 reaches a critical value related to the entropy. However, in cases where the physical parameter $s$ is bounded, its variance must remain finite, as has been previously noted by Klein and Tisza [30]. The experimental evidence for the divergence relates to opalescence near the critical point of the gas-liquid transition [31], under conditions where Eqn 1 is still valid. When the fluctuations become at all large, the probe beam is multiply scattered and the fluctuation amplitude can no longer be determined. Clearly, the density fluctuations are bounded below by zero and above by the density at which the molecules are close-packed in the Kitaigorodskii sense [32]. They can become large but cannot diverge.

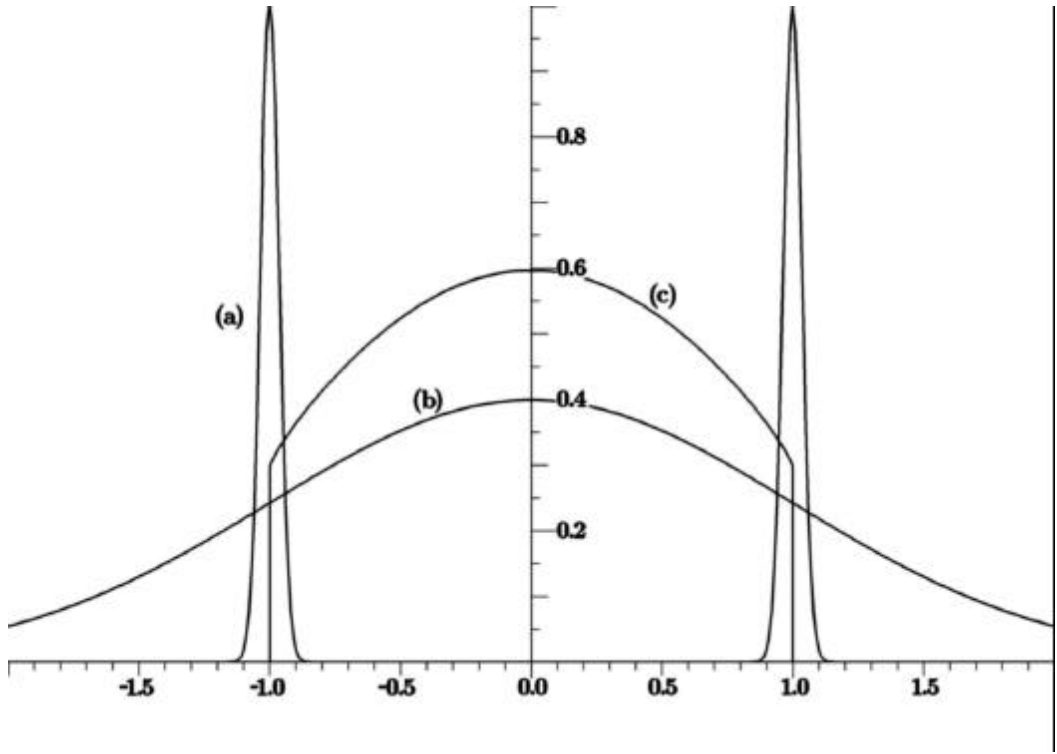

Figure 1. A comparison of probability distribution functions for the case of an isolated spin:
- true (delta functions shown stylized for clarity)
- effective Gaussian with the same mean and standard deviation
- exp (-Φ/kT)

## 5.  DISCUSSION

The analysis presented here has nothing to say about the behavior of a thermodynamic system with changes of temperature, pressure or magnetic field, and *a fortiori* nothing about its critical behavior at a phase transition. However, it is most certainly relevant to the system structure at molecular-scale resolution. When the Landau theory successfully predicts equilibrium phase behavior, the present analysis justifies the use of its free-energy functional in inquiries into the spatial correlations of the order within each individual phase. This order has experimental implications through its effect on scattering of X-rays, electrons and neutrons.

The molecular Hamiltonian is not amenable to mathematical analysis, while Kadanoff's 'coarse-graining' procedure, which has been so successful in allowing the approximate evaluation of the potentials, inherently destroys molecular-scale structure and does not assist the present investigation. The Guggenheim definition of the free-energy functional proves both relevant and amenable. Just as for the free energy itself, calculation of this functional must be undertaken using the modern armory of statistical thermodynamic techniques.

The present paper has resolved two issues important for the present goal. Firstly, under conditions far from a critical point, Eqn 1 provides an approximation to the fluctuation statistics adequate for present purposes when $\Phi$ is taken to be the Guggenheim functional. However Eqn 1 is not an exact representation of the statistics, giving correctly only the first and second moments. The implicit divergence in the vicinity of a critical point is an artifact of the approximation. In physical systems, the fluctuations remain bounded.

Secondly, the minimum of the Guggenheim free-energy functional is equal to the thermodynamic free energy, with no term correcting for fluctuations.

## ACKNOWLEDGEMENTS

The author would like to express his gratitude to Dr. V.M. Kaganer for useful discussions.